\documentclass[sigconf]{acmart}

\usepackage{kotex}
\usepackage{caption}
\usepackage{subcaption}

\usepackage{graphicx} 
\usepackage[dvipsnames]{xcolor}
\usepackage[utf8]{inputenc}
\usepackage{kotex}
\usepackage{enumitem} 
\usepackage{subcaption}
\usepackage{multirow}
\usepackage{amsmath} 
\usepackage{booktabs} 
\usepackage{threeparttable} 
\usepackage{array} 
\usepackage{circledsteps}
\usepackage{verbatim} 
\usepackage{ulem}

\newif\ifdebug
\newif\ifdiff
\newif\iffinaldiff

\debugtrue
\difftrue

  {\list{}{\leftmargin=#1\rightmargin=#1}\item[]}%
  {\endlist}

\definecolor{bluebox_fill}{cmyk}{0.27, 0.01, 0.01, 0}
\definecolor{bluebox_border}{cmyk}{0.79, 0.44, 0.02, 0.01}

\newcounter{unknowncounter}
\newcounter{commentcounter}
\newcounter{needchartcounter}
\newcounter{needfigcounter}
\newcounter{needtablecounter}
\newcounter{needcitecounter}

\newcommand{\drop}[1]{\ifdiff{\color{pink}{}}\else{}\fi}

\copyrightyear{2025}
\acmYear{2025}
\setcopyright{rightsretained}
\acmConference[UIST Adjunct '25]{The 38th Annual ACM Symposium on User Interface Software and Technology}{September 28-October 1, 2025}{Busan, Republic of Korea}
\acmBooktitle{The 38th Annual ACM Symposium on User Interface Software and Technology (UIST Adjunct '25), September 28-October 1, 2025, Busan, Republic of Korea}\acmDOI{10.1145/3746058.3758440}
\acmISBN{979-8-4007-2036-9/2025/09}

\begin{document}
\title{Chameleon: A Surface-Anchored Smartphone AR Prototype with Visually Blended Mobile Display}

\author{Seungwon Yang}
\email{sw.yang@postech.ac.kr}
\affiliation{%
  \department{Dept. of CSE}
  \institution{POSTECH}
  \state{Pohang}
  \country{South Korea}
}

\author{Suwon Yoon}
\email{suwon.yoon@postech.ac.kr}
\affiliation{%
  \department{Dept. of CSE}
  \institution{POSTECH}
  \state{Pohang}
  \country{South Korea}
}

\author{Jeongwon Choi}
\email{choijw@postech.ac.kr}
\affiliation{%
  \department{Dept. of CSE}
  \institution{POSTECH}
  \state{Pohang}
  \country{South Korea}
}

\author{Inseok Hwang}
\email{i.hwang@postech.ac.kr}
\affiliation{%
  \department{Dept. of CSE}
  \institution{POSTECH}
  \state{Pohang}
  \country{South Korea}
}

\renewcommand{\shortauthors}{Yang et al.}

\begin{abstract}
Augmented reality (AR) is often realized through head-mounted displays, offering immersive but egocentric experiences. While smartphone-based AR is more accessible, it remains limited to handheld, single-user interaction. We introduce Chameleon, a prototype AR system that transforms smartphones into surface-anchored displays for co-located use. When placed flat, the phone creates a transparency illusion and anchors digital content visible to multiple users. Chameleon supports natural repositioning on the surface without external hardware by combining two techniques: (1) Background Acquisition uses opportunistic sensing and language model-assisted pattern generation to blend with surrounding surfaces, and (2) Real-Time Position Tracking augments inertial sensing to maintain spatial stability. This work shows how lightweight sensing can support casual, collaborative AR experiences using existing devices.
\end{abstract}



\begin{teaserfigure}
    \centering
    \includegraphics[width=\linewidth]{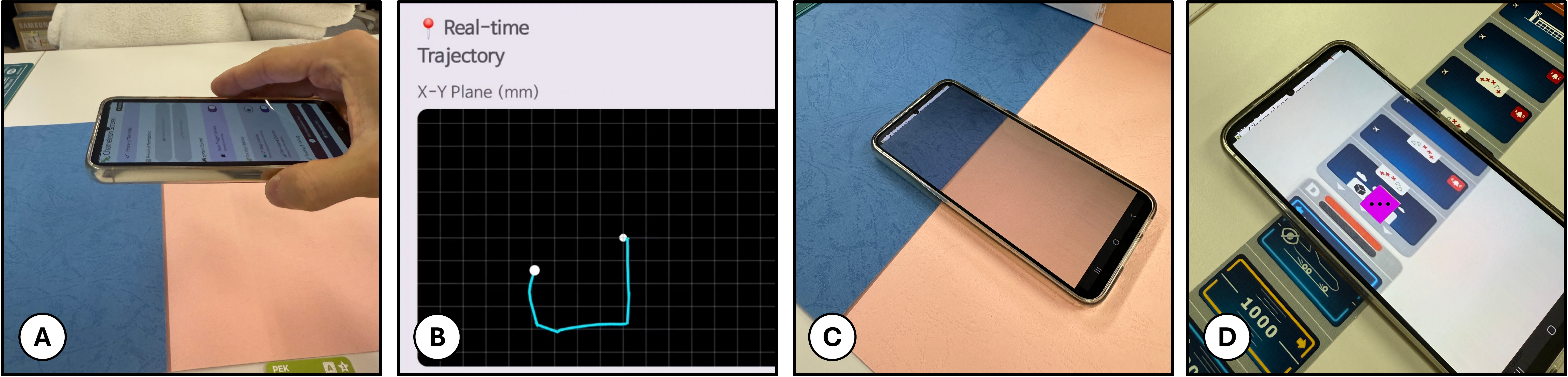}
    \caption{Overview of the Chameleon system. (A) Background Acquisition. (B) Real-Time Position Tracking. (C) Visually Blended Mobile Display. (D) Embodied Dice Roll in Physical Games}
    \label{fig:teaser}
    \Description{(A) The hand holds a smartphone horizontally. The screen shows several toggle options. The smartphone is about 15 cm above the surface. On the surface, there are two papers: blue on the left and pink on the right. The smartphone is pointing at the border between the papers. (B) A real-time trajectory screen is shown. The X-Y plane is presented in a millimeter scale. The black grid shows a U-shaped trajectory in a light blue line. (C) The smartphone is laid on the surface, on the border between the blue and pink papers. The smartphone screen also shows the blue and pink papers, as if the smartphone were transparent. (D) The smartphone is laid on the board game cards. The smartphone screen shows the white surface of the table and part of the board game cards beneath it, making it seem transparent. A virtual die is placed on the screen. The visible side of the dice is pink with the number three.}
\end{teaserfigure}

\maketitle

\section{Introduction}
\label{sec:intro}

Augmented Reality (AR) is often envisioned through head-mounted displays (HMDs), where digital content is overlaid onto the user’s first-person perspective. These systems offer immersive experiences but tend to be egocentric, requiring dedicated hardware and limiting interactions to the individual wearer. While smartphone-based AR is more accessible and requires no special equipment~\cite{cho2022touchvr}, it remains inherently handheld and personal. The display is visible only to the person holding the device, making shared interaction difficult. As a result, AR is often constrained by either hardware complexity or limited engagement modes.




To address this, we developed Chameleon, a prototype AR system that transforms smartphones into surface-anchored displays. When placed flat, Chameleon creates transparency illusions while anchoring digital content visible to multiple users, enabling co-located interaction without specialized equipment.

Prior research has explored various approaches to smartphone-based surface interaction. Custom mobile devices with see-through displays~\cite{pearson2017chameleon} support interaction with documents beneath the screen~\cite{hincapie2014car}, while other systems employ a smartphone's front-facing camera and external markers for position determination on documents to create the transparency illusion~\cite{pucihar2014poster}. Chameleon differs through its combination of opportunistic sensing and inertial-based tracking, achieving general-purpose, marker-free operation without being limited to specific surface types or domains.

Specifically, Chameleon contributes two technical innovations. Background Acquisition employs opportunistic surface capture when users lift their devices, with explorations into camera-in-contact sensing~\cite{yeo2017specam} and LLMs for intelligent scene blending. Real-Time Position Tracking relies on inertial sensing, investigating magnetic correction, and drift-tolerant algorithms for stability during collaborative use.

We present Chameleon as an investigation of how these sensing and tracking strategies enable practical shared AR within existing hardware constraints. The resulting system achieves low-cost deployment, low-effort setup through automatic capture, and low-constraint usage that tolerates natural device movements. This work demonstrates that accessible collaborative AR may be realized through creative applications of smartphone capabilities rather than new devices.

\section{Design and Implementation}
\label{sec:design}

Chameleon is composed of two key components: (i) a calibration pipeline that captures and prepares the background image for camouflage rendering, and (ii) a real-time tracking and rendering engine that maintains visual alignment as the smartphone is moved across the surface. 

\subsection{Background Acquisition}
To achieve the illusion of transparency~\cite{wigdor2007lucid}, Chameleon first requires an image of the surface beneath the smartphone. Rather than asking users to manually trigger a capture, the system passively monitors the device orientation and motion. When the phone is held parallel to the surface for a short period, the system automatically captures an image of the surface from above.

This image serves as the reference background. After capturing, users can place the phone flat on the surface. The system uses the captured image to render the portion of the surface that is occluded by the phone screen, effectively camouflaging the device against its environment.

\subsection{Real-Time Position Tracking and Rendering}

Once the phone is placed on the surface, Chameleon begins to render the occluded region in real-time, updating the view based on the phone's position and orientation. To track how the phone moves across the surface, Chameleon uses IMU-based dead reckoning to estimate the device's displacement and rotation relative to its initial placement. 

To address the cumulative drift and sensor bias that naturally accumulate in dead reckoning, we implement an extended Kalman filter to maintain a stable estimate of position. 
To reduce vertical error, the system uses downward-facing camera images to confirm surface contact, allowing motion to be simplified to 2D tracking.

To further contain drift, we apply a zero velocity update whenever the phone is detected to be stationary. During these periods, displacement updates are halted to prevent drift from accumulating unnecessarily.

\section{Applications}
\label{sec:applications}



Chameleon enables a class of augmented reality experiences that blend into everyday surfaces without requiring specialized hardware. Because it preserves the visual context of the underlying surface, it can be naturally integrated into analog-to-digital workflows. Although the system operates solely on a smartphone, it supports key features of AR such as spatial alignment, contextual blending, and co-located interaction. These qualities make it well suited for everyday, shared environments. We present three example use cases:

\textit{Embedded Dice Roll in Physical Games. }
Chameleon creates a dice-rolling zone that appears built into a game board, allowing digital interactions without disrupting the flow of physical play~\cite{cho2023ai}.

\textit{Camouflaged Paperweight. }
Placed over documents, the phone reveals content beneath while acting as a movable paperweight, supporting shared reading without visual obstruction.

\textit{Augmented Annotation Lens. }
As the phone moves across a printed diagram, it reveals aligned annotations or translations, enabling layered discussion without external displays.


\section{Design Implications and Conclusion}
\label{sec:discussion}

Chameleon explores how AR experiences can emerge from minimal hardware and blend naturally into existing physical environments. By anchoring interaction on shared surfaces rather than individual viewpoints, it shifts the focus from egocentric AR to situated and co-located use. This presents a complementary design space to immersive AR, one that prioritizes simplicity, portability, and shared presence over full-field augmentation.

Our approach shows that visual blending and spatial stability can support meaningful interaction, even on a small, handheld screen. It invites new design considerations around visibility, analog-digital integration, and casual multi-user interaction in everyday settings.

Moving forward, we aim to expand Chameleon’s capabilities through two lines of technical exploration. First, we are investigating \textit{camera-in-contact sensing}, where the device recognizes color and texture directly from the surface beneath it. Assuming access to a larger-scale photo of the surrounding environment, the system infers the surface color from the sampled color and prompts a large language model (LLM) to identify the most plausible match within the space. The LLM then generates a coherent surface pattern that blends with the perceived background. While not pixel-accurate, this process leverages minimal visual input to produce textures that match both visually and contextually.

Second, we explore \textit{real-time position tracking} by augmenting inertial sensing with the smartphone’s built-in magnetic sensors~\cite{kim2023proxifit}. Because IMU-based tracking is prone to drift, we are testing lightweight correction methods that leverage ambient magnetic field data from easily found household magnets or magnetic properties of surrounding surfaces to improve spatial stability without requiring external infrastructure.

\textit{Limitations.}
The current system uses IMU-based dead reckoning, which is prone to drift over time. The bezel of the phone can interfere with the illusion of full transparency. Because interaction occurs on a flat surface without adapting to the user’s viewpoint, the display can appear visually flat. The limited screen size of smartphones also constrains the scale of interaction.

\begin{acks}
This research was supported by the National Research Foundation of Korea (NRF) grant funded by the Korea government (MSIT) (RS-2025-00553946). This research was also partly supported by Seoul R\&BD Program (SP240008) through the Seoul Business Agency (SBA) funded by The Seoul Metropolitan Government and a grant (RS-2025-00564342) from the Korea Institute for Advancement of Technology (KIAT) funded by the Korea government (MOTIE).
\end{acks}

\bibliographystyle{ACM-Reference-Format}
\bibliography{references}

\end{document}